# Twist-controlled resonant tunnelling in graphene – boron nitride – graphene heterostructures

A. Mishchenko[1], J. S. Tu[2], Y. Cao[2], R. V. Gorbachev[2], J. R. Wallbank[3], M.T. Greenaway[4], V. E. Morozov[1], S. V. Morozov[5], M. J. Zhu[1], S. L. Wong[1], F. Withers[1], C. R. Woods[1], Y.-J. Kim[2,6], K. Watanabe[7], T. Taniguchi[7], E. E. Vdovin[4,5], O. Makarovsky[4], T. M. Fromhold[4], V. I. Fal'ko[3], A. K. Geim[1,2], L. Eaves[1,4], K. S. Novoselov[1]

[1]School of Physics & Astronomy, University of Manchester, Oxford Road, Manchester, M13 9PL, UK

[2]Centre for Mesoscience & Nanotechnology, University of Manchester, Manchester, M13 9PL, UK

[3]Physics Department, Lancaster University, LA1 4YB, UK

[4]School of Physics and Astronomy, University of Nottingham, Nottingham NG7 2RD, UK

[5]Institute of Microelectronics Technology and High Purity Materials, Russian Academy of Sciences, Chernogolovka, 142432, Russia

[6]Department of Chemistry, Seoul National University, Seoul 151-747, Korea

[7]National Institute for Materials Science, 1-1 Namiki, Tsukuba 305-0044, Japan

**Recent developments in the technology of van der Waals heterostructures[1,2] made from two-dimensional atomic crystals[3,4] have already led to the observation of new physical phenomena, such as the metal-insulator transition[5] and Coulomb drag[6,7], and to the realisation of functional devices, such as tunnel diodes[8,9], tunnel transistors[10,11] and photovoltaic sensors[12]. An unprecedented degree of control of the electronic properties is available not only by means of the selection of materials in the stack[13] but also through the additional fine-tuning achievable by adjusting the built-in strain and relative orientation of the component layers[14-18]. Here we demonstrate how careful alignment of the crystallographic orientation of two graphene electrodes, separated by a layer of hexagonal boron nitride (hBN) in a transistor device, can achieve resonant tunnelling with conservation of electron energy, momentum and, potentially, chirality. We show how the resonance peak and negative differential conductance in the device characteristics induces a tuneable radio-frequency oscillatory current which has potential for future high frequency technology.**

The growing catalogue of 2D crystals allows us to construct increasingly complex van der Waals heterostructures[8-12,19,20]. The combination of a hexagonal boron nitride barrier layer sandwiched between two graphene electrodes is particularly attractive[8,9,21] due to the exceptional crystalline quality and the small lattice mismatch of these two materials. For example, by utilising a third (gate) electrode, it has recently proved possible to make a novel type of field-effect transistor in which tunnelling between the two graphene electrodes is controlled by gate voltage[10,22]. In the prototype versions of these devices, the crystalline lattices of the component layers were not intentionally aligned[8,9,21], which meant that tunnelling between the two graphene electrodes required a defect-assisted momentum transfer, so that the tunnelling was not resonant.

Here, we report on a new series of tunnel transistors in which the crystal lattices of the two graphene layers are intentionally aligned to a high degree of precision during the fabrication procedure. Our measurements and theoretical modelling of the device characteristics reveal that the current flow is dominated by tunnel transitions in which both energy and in-plane momentum are conserved. The resonant conditions exist in a narrow range of bias voltages, and result in a resonant peak in the current-voltage characteristics, leading to strong negative differential conductance (NDC). In the NDC region, our devices generate radio frequency oscillations when connected to a simple *LC* circuit. This proof-of-principle experiment points the way towards new applications for graphene-based resonant tunnelling devices in high-frequency electronics. In addition, the development of a fabrication procedure that aligns precisely the crystalline lattices of the component layers of our devices eliminates the variability in the device characteristics which arises when the graphene electrodes are randomly oriented, thus advancing the prospects for future manufacturable technology applications. Furthermore, control of the misalignment angle and adjustment of the thickness and/or composition of the tunnel barrier could be used to fine-tune the device characteristics.

A schematic diagram of our transistor is shown in Fig. 1a. The heterostructure is made by means of a standard dry-transfer procedure of mechanically-exfoliated graphene and hBN layers[5,23], with the important additional step that the lattices of the top (*T*) and bottom (*B*) graphene flakes are aligned to within 2° of each other. We used mechanically torn graphene flakes[24] with well-defined facets and were able to distinguish between the armchair and zig-zag edges by comparing the intensity of the Raman D peak from the edges[25-27]. This allowed us to know the crystallographic orientation of both top and bottom graphene, thus achieving high level of alignment (see Supplementary Information for details). An independent prove of the crystallographic alignment between the two graphene electrodes comes from the measurements of the broadening of the Raman 2D peak for the two graphene flakes. Such broadening serves as a measure of the rotation angle between graphene and underlying hBN[13,28], and allows one to calculate the relative angle between the crystallographic directions for the two graphene electrodes (see Supplementary Information for details). In order to improve its electronic quality, the bottom graphene electrode was placed on a thick layer of hBN overlaying the $SiO_2$/Si substrate, with the heavily doped Si wafer acting as a back gate[29]. The two graphene layers were independently contacted with Cr/Au metallization.

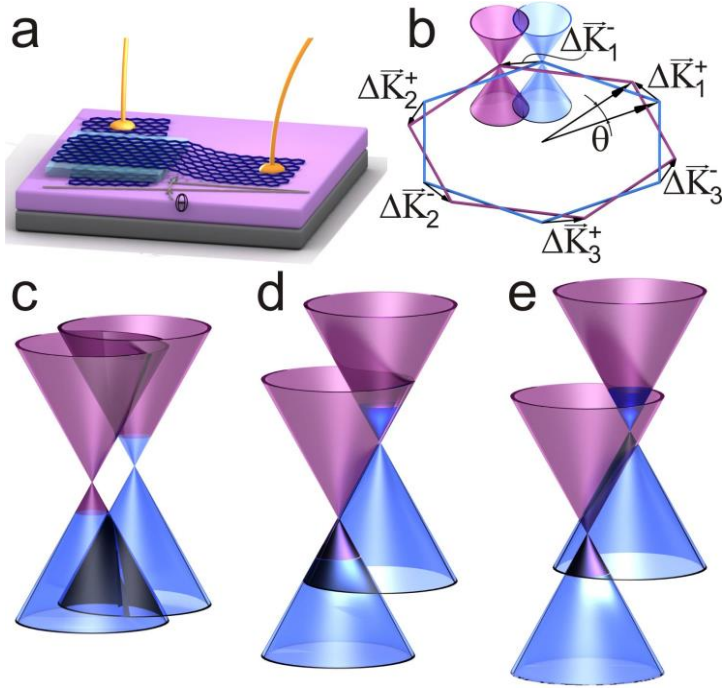

**Figure 1:** Schematic representation of our device and its band structure. **a)** Device schematics with the angle $\theta$ between two graphene layers (separated by a BN tunnel barrier shown in light blue) exaggerated, **b)** A rotation by $\theta$ of the two graphene layers in real space corresponds to the momentum shift $\Delta K_i^{\pm}$ between two Dirac points. Diagrams **c-e** represent the relative alignment between top (left cones) and bottom (right cones) graphene Dirac points; the boundary between magenta (empty states) and blue (filled states) colours marks the Fermi level.

We chose the thickness of hBN tunnel barriers to be 4 monolayers, which allows us to work in a comfortable range of current densities and bias voltages. Note, that the current density can be dramatically increased (4-5 orders of magnitude)[8,9] if thinner (2-3 layers) hBN or a material with lower tunnel barrier (like $WS_2$) is utilised[10,11,19].

Figure 2a shows the dependence of the current density, $J$, measured at 2K as a function of the bias voltage, $V_b$, applied between the two graphene electrodes for three different values of the gate voltage, $V_g$, applied between the lower graphene electrode and the doped Si substrate. Note, that the value of the current density in these devices is an order of magnitude larger than in previous devices with the same barrier thickness, but with misoriented electrodes[10,21]. We observe a strong peak in $J(V_b)$, followed by a region of NDC, both of which persist up to room temperature. We attribute this peak to resonant tunnelling of carriers between the two graphene electrodes (with momentum conservation). In order to display in more detail the key features of the device characteristics, Fig. 2b and c present colour scale contour maps of the dependence of the differential conductance ($dI/dV_b$) and $d^2I/dV_b^2$ on $V_b$ and $V_g$ (here $I$ is current). In Fig. 2b the regions of the NDC are shown as blue areas. Furthermore, weaker resonances can be seen as a transition from pink to red colours.

In order to explain the physics of the electron tunnelling in these devices, we used a theoretical model[30-33] which takes into account the unique band structure of graphene, the physics of the eigenstates of the massless Dirac fermions in the graphene layers and the effect of temperature on the device characteristics. The results of this model, shown in Fig. 2d-f, reproduce the measured device characteristics. Here, we focus on the case of a small angular misalignment, θ, of the two graphene

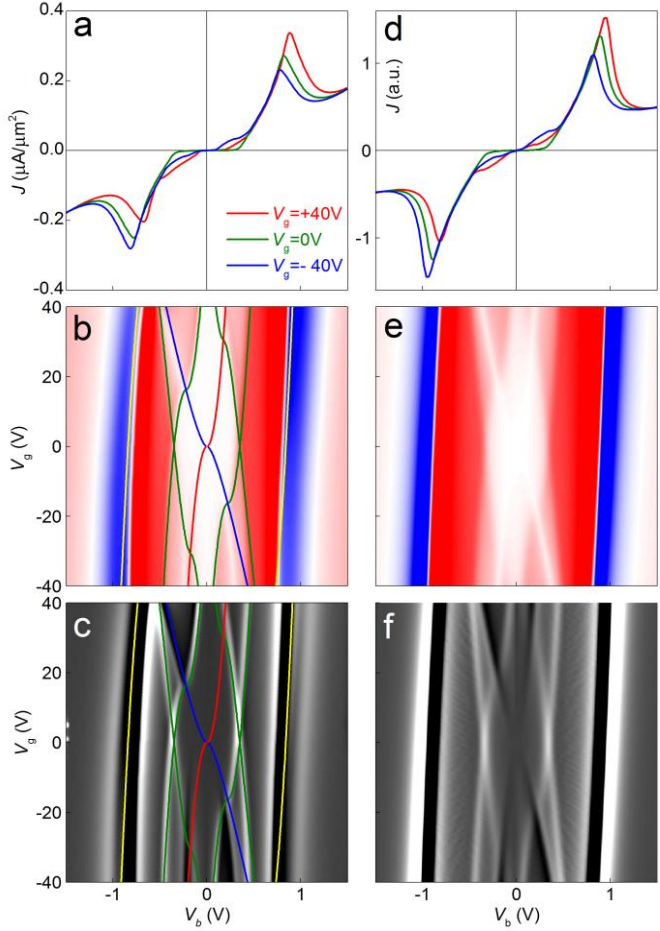

**Figure 2.** The device tunnelling characteristics at 2K. Panels (a)-(c): experiment; (d)-(f): theoretical simulations. a,d) Current density-voltage curves at different $V_g$. b,e) Conductivity $dI/dV$ plots as a function of $V_b$ and $V_g$ (colour scale is blue to white to red: -0.7μS to 0μS to 0.7μS). c,f) $|d^2I/dV^2|$ plots as a function of $V_b$ and $V_g$. The dashed lines in b) and c) are theoretical simulations and correspond to different resonant conditions. Red and blue lines correspond to events when the Fermi level in one of the graphene layers passes through the Dirac point. The green lines correspond to the event depicted in Fig. 1c, and the yellow line – to the event depicted in Fig. 1d. Device characteristics: active area of ≈8 μm², hBN barrier thickness is 1.4nm (4 layers), $\theta \approx 1.8°$.

lattices – see Fig. 1a, which corresponds to a rotation of the two graphene Brillouin zones in *k*-space, see Fig. 1b. In particular, the neutrality points at the six *K*-points of the Brillouin zone, $\hbar K_i^\pm$ (where i=1,2,3 identify equivalent corners, ± distinguish between *K* and *K'* valleys and $\hbar$ is the reduced Plank constant) are displaced by a wavevectors $\hbar \Delta K_i^\pm = l_z \times \theta K_i^\pm$.

The resulting intersection of the Dirac cones which visualise the conditions for resonant tunnelling of electrons between two layers are shown in Fig. 1c-e for characteristic regimes of bias voltage. They display the energy shift of both Dirac cones and Fermi levels as the carrier concentrations in the two graphene layers change due to changes in the bias and gate voltages. Fig. 1c and, especially, Fig. 1d illustrate how conditions for resonant tunnelling can be satisfied. The case shown in Fig. 1d is of particular importance, since the tunnelling current is maximized when the momentum difference, $\Delta K$, is compensated by changing electrostatically the energy of the two Dirac cones by an amount $\pm \hbar v_F \Delta K$ (valid for small θ, here $v_F$ is the graphene Fermi velocity). In this case the conical dispersions in the two layers intersect along a straight line and a large fraction of the states along that line are occupied in one layer and empty in the other layer, thus facilitating a large resonant tunnel current. Fig. 1e demonstrates the case at yet higher bias voltage between the top and bottom graphene layers, where the in-plane momentum is conserved only for a small number of states (far away from the Fermi levels of the two graphene electrodes), thus, leading to the reduction in the current. We now consider the detailed features of the measured differential conductance map, Fig.2c. The red and blue dashed lines in Fig. 2c are theoretical simulations and correspond to the situation when Fermi levels of the top and bottom graphene layers pass through Dirac points. These lines are universal features for all graphene-to-graphene tunnelling systems regardless of their relative alignment, and can be obtained using a simple electrostatic model (see Supplementary Information). The density of states (*DoS*) of graphene close to the Dirac point varies linearly with chemical potential, i.e., $DoS \propto |\mu|$. Thus, the tunnel conductance decreases when the graphene Fermi level crosses the Dirac point due to the vanishing *DoS* at this point. The small but finite conductance at the Dirac point is due to charge inhomogeneity, which results in *DoS* smearing at the Dirac point and remains non-zero[34].

The set of green lines in Fig. 2b and c trace the low bias and relatively weak resonance in the device characteristics when the chemical potential of one of the graphene layers reaches the point where the two Dirac cones begin to overlap (Fig. 1c). These four lines satisfy the condition $\mu_{T,B} = (\Delta\varphi \pm \hbar v_F \Delta K)/2$. The intersection of these lines at $V_g = 0$ provides a good measure of the momentum mismatch $\Delta K = V_b/\hbar v_F$, and therefore gives an accurate estimate of the misalignment angle, $\theta$. For example, for the device shown in Fig. 2, we estimate that the angle $\theta = 1.8°$. It is interesting to note here the geometry of the intersection of the two Dirac cones under the conditions shown in Fig. 2: they form hyperbolic solutions. The wavevectors of the Dirac plane waves lie on three hyperbolae obtained by 120° rotations of these hyperbolic solutions (see Supplementary Information). The solutions remain hyperbolic for $|\Delta\varphi| < \hbar v_F \Delta K$. This low bias resonance does not lead to NDC, but it does give rise to a significant increase in conductance, as shown in Fig. 2b.

The voltage dependence of the main resonance peak is shown as yellow lines in Fig. 2b and c. In this case, the Dirac cones are shifted by $\Delta\varphi = \pm\hbar v_F \Delta K$ so that the intersection of the cones is a straight line (i.e. the wavevectors of the Dirac plane waves lie on straight lines), see Fig. 1d. In this situation, momentum is conserved for tunnelling electrons at all energies between $\mu_T$ and $\mu_B$, thus giving rise to a strong peak in the current density at resonance. When the Dirac cones are displaced further beyond this resonant condition, that is for $|\Delta\varphi| > \hbar v_F \Delta K$, the curve of intersection of the two Dirac cones becomes an ellipse (Fig. 1e). As a result, the wavevectors lie on three ellipses (again, obtained by 120° rotations of elliptic solutions - see Supplementary Information). In contrast to the hyperbolic and linear solutions, the elliptic solutions are bound – only wavevectors limited by these ellipses can contribute to current density. The reduction of the current once the Dirac cones are shifted further off-resonance is the physical mechanism which gives rise to the negative differential conductance region beyond the resonant peak.

The main resonant peak, which would be Dirac-delta like in the absence of broadening, has a finite width due to the presence of short-range scatterers, charge inhomogeneity in the graphene layers (electron-hole puddles) or orientational disorder between two graphene layers because of bubble formation[1]. Furthermore, since this mechanism of resonant tunnelling relies on momentum conservation, the position of the resonant peak and the peak-to-valley ratio are only weakly dependent on temperature. This mechanism for NDC is only possible in the graphene-graphene tunnelling system; for example, if one or both electrodes are replaced with bilayer graphene, then, due to the parabolic dispersion relation, the extended linear intersection is no longer possible. Note that in the modelled *J-V*$_b$ characteristics (Fig 2d-f) we take into account the chirality of electrons (the momentum-dependent phase shift of the wavefunction on the two sublattices in graphene, see Supplementary information). In the absence of strain, however, the results stay qualitatively the same.

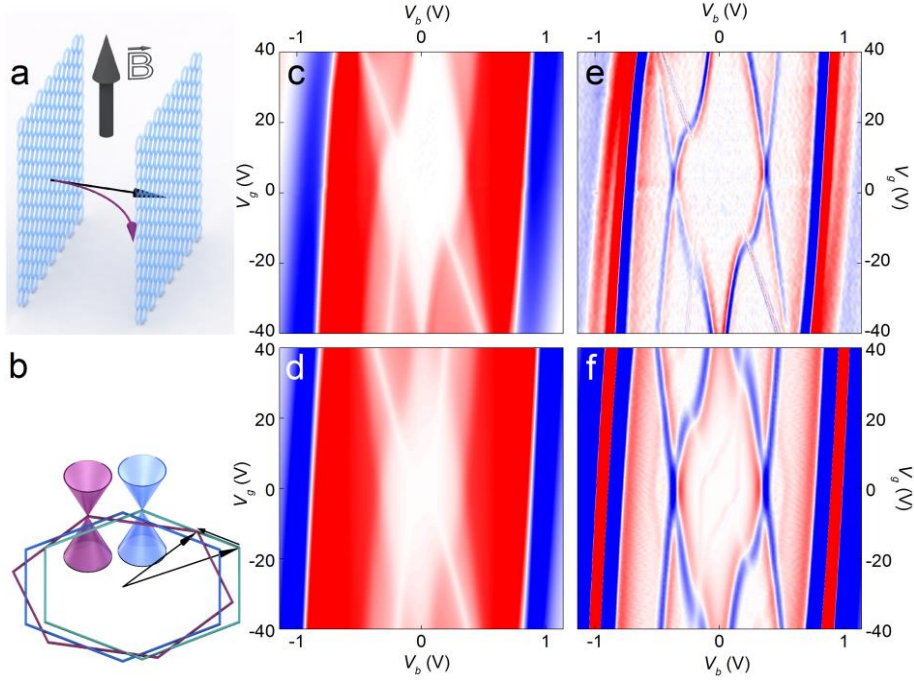

**Figure 3.** Effect of in-plane magnetic field on resonant tunnelling, same device as in Figure 2. Panels c) and e) are experimental data, panels d) and f) – theoretical modelling. a) Trajectories of the charged quasiparticles in zero (black arrow) and finite (purple arrow) magnetic field due to Lorentz force. b) The resulting shift of Fermi surface due to an in-plane magnetic field, exaggerated for clarity, c,d) *dI/dV* maps measured with 15T in-plane magnetic field applied, e,f) difference between *dI/dV* maps with and without in-plane *B* field.

In order to confirm our proposed mechanism of resonant tunnelling we performed additional measurements in which a magnetic field, $B_\parallel$, was applied parallel to the graphene layers, i.e. perpendicular to the tunnel current. Classically, the electron tunnelling between two 2D electrodes through a barrier of thickness *d* will acquire an additional in-plane momentum $ed\bm{l_z} \times \bm{B}_\parallel$, due to the action of the Lorentz force[35,36], which modifies the resonance. Depending on the orientation of the magnetic field with respect to crystallographic directions of the two graphene layers, the magnitude of the wavevectors,

$$\hbar\Delta\bm{K}_i^\pm = \bm{l_z} \times \left[\theta\bm{K}_i^\pm + ed\bm{B}_\parallel\right]$$

differs for each of the six Dirac cones at the corners of Brillouin zone.

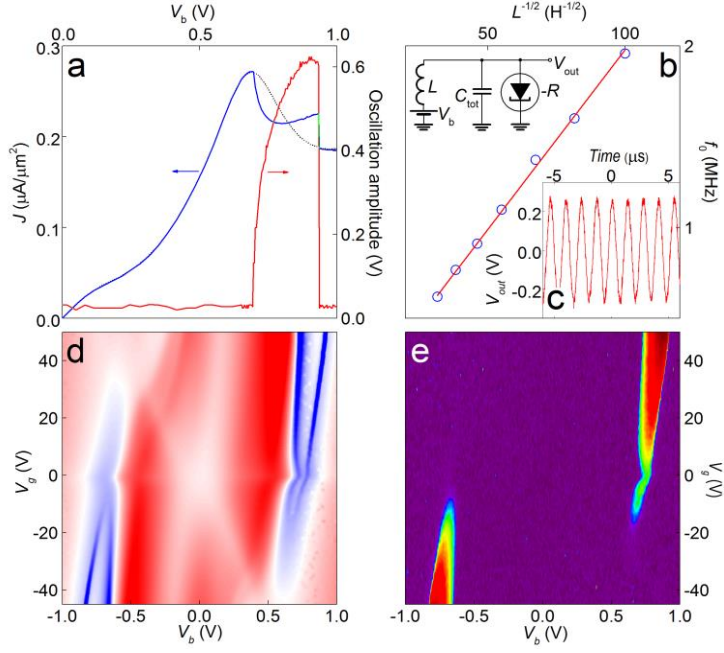

**Figure 4.** Radio-frequency oscillator based on resonant tunnelling transistor, $T=300$K. a) $J(V_b)$ characteristics with (blue, solid, $L=1$mH) and without (black, dashed, $L=0$H) external $LC$ circuit. The red curve shows the $V_b$-dependence of the peak-to-peak amplitude of the oscillations. b) Resonant frequency of the oscillator (schematics shown on the top left inset) vs. inductance of the $LC$ circuit. For this device the total capacitance of the assembly was estimated from a simple circuit model (see top left inset) to be 65pF. c) $V_{out}$ as a function of time for $L=1$mH. d) d$I$/d$V_b$ map measured with a 330μH inductance (red to white to blue, -0.7μS to 0μS to 0.7μS) and e) corresponding amplitude map (red to yellow to violet, 0.5V to 0.26V to 0.03V). Device structure: active area of ≈120 μm², hBN spacer =1.4nm (4 layers), $\theta \approx 0.9°$.

Figure 3 presents the measured (c) and calculated (d) tunnel conductance maps in the presence of a strong in-plane magnetic field, $B_\parallel$. In Fig. 3e, the measured contour maps at zero field (Fig. 2b) and with (Fig. 3c) magnetic field ($B_\parallel = 15$ T) are subtracted from each other. This "difference map" reveals that the bias positions of the resonances, which correspond to various momentum conservation conditions, are shifted significantly by the in-plane field. At the same time, the position of the $\mu_T = 0$ and $\mu_B = 0$ resonances (which are insensitive to momentum conservation) in the d$I$/d$V_b$ contour maps are almost unchanged. The fact that the in-plane magnetic field affects only the resonances where momentum conservation processes are involved, rules out the possible artefacts, such as a small perpendicular component of magnetic field.

One of the celebrated applications of devices with NDC is their use as high frequency oscillators, which are typically constructed by connecting an external resonant circuit to the device. Previously, it has been demonstrated that graphene/hBN/graphene devices should not exhibit instabilities which would result in intrinsic oscillations[37]. In conventional double barrier resonant tunnelling devices (DBRTDs), the build-up of charge in the quantum well leads to a delay of the current with respect to the voltage which can be represented in terms of an inductance in the equivalent circuit of the device[38-40]. This effective inductance, which is an important feature in the operation of DBRTD oscillators, is absent in our single barrier devices[37]. Thus, as a proof-of-principle we have built such an oscillator by adding an inductance in series with our resonant tunnelling device, while utilising the intrinsic and parasitic capacitance ($C_{tot}$) as a capacitance of $LC$ circuit, see Figure 4b. When the bias and gate voltages are tuned to the NDC region, the device undergoes stable sine-wave oscillations, see

Fig. 4a,c. The oscillation frequency can be tuned by varying the parameters of the external circuit, see Fig. 4b.

The operation of an NDC-based *LC* resonator can be understood as follows. Once excited, the *LC* circuit produces damped oscillations which rapidly decay to zero. This is mainly due to the internal dissipative resistance, *R*, of the resonator and other losses. When the tunnel transistor operates in the NDC region, its negative resistance cancels the internal lossy resistance, thus supporting continuous stable oscillations at the resonant frequency of the *LC* circuit. Equivalently, in the NDC region the tunnel transistor provides amplification (positive feedback), and stable oscillations occur if the gain exceeds unity. Interestingly, the shape of the $J$-$V_b$ curve changes in the region of stable oscillations, as compared with the case without the *LC* circuit, see Figure 4a (likely, due to the change in the asymmetric rectification of RF oscillations in the strongly nonlinear $J$-$V_b$ region).

**Conclusions**

By aligning the crystallographic orientation of the two graphene layers in a graphene-hBN-graphene heterostructure, we have demonstrated that resonant tunnelling with both energy and momentum conservation can be achieved. This results in strong NDC which persists up to room temperature. The bias position of the resonance can be controlled by the relative orientation between the two graphene crystalline lattices and by external magnetic field. Our tunnel diodes produce stable oscillations in the MHz frequency range, limited mainly by the parasitic capacitance between the contact pads of our devices and the underlying Si gate. Much higher frequencies could be reached by reducing this parasitic capacitance and that of the external circuit. Even higher frequencies could also be achieved by fabricating a device in a slot antenna configuration, where the slot acts as a resonator with resonance frequency determined by the geometry of the slot. Moreover, our tunnel devices are free of the fundamental limitation intrinsic to conventional double barrier resonant tunnelling devices, namely the relatively long carrier dwell time (~ps) in the quantum well as compared to the time to transit the barrier (~fs). This suggests that such tunnel circuits can be potentially scaled to operate in the THz regime.

Correspondence and requests for materials should be addressed to LE (Laurence.Eaves@nottingham.ac.uk ) and KSN (kostya@manchester.ac.uk ).

This work was supported by the European Research Council, EC-FET European Graphene Flagship, Engineering and Physical Sciences Research Council (UK), the Leverhulme Trust (UK), the Royal Society, U.S. Office of Naval Research, U.S. Air Force Office of Scientific Research, U.S. Army Research Office and RS-RFBR, grants number 14-02-00792 and 13-02-92612 (Russian Federation). Y.-J.K. was supported by the Global Research Lab Program (2011-0021972) through the National Research Foundation of Korea funded by the Ministry of Science, ICT & Future, Korea.